\documentclass[aps,pre,twocolumn,superscriptaddress]{revtex4-1}
\usepackage{graphicx}
\usepackage[rgb,dvipsnames]{xcolor}
\usepackage{amsmath}
\usepackage{amsthm, amssymb}
\usepackage{epsfig}
\usepackage{hyperref}
\usepackage{bm}

\usepackage{natbib}

\usepackage[normalem]{ulem}

\usepackage{placeins}
\usepackage[symbol]{footmisc}

\begin{document}

\title[Flags, Landscapes and Signaling]{Flags, Landscapes and Signaling:~ Contact-mediated inter-cellular interactions enable plasticity in fate determination driven by positional information}

\author{Chandrashekar Kuyyamudi}
\affiliation{The Institute of Mathematical Sciences, CIT Campus, Taramani, Chennai 600113, India}
\affiliation{Homi Bhabha National Institute, Anushaktinagar, Mumbai 400 094, India}
\author{Shakti N. Menon}
\affiliation{The Institute of Mathematical Sciences, CIT Campus, Taramani, Chennai 600113, India}
\author{Sitabhra Sinha}
\affiliation{The Institute of Mathematical Sciences, CIT Campus, Taramani, Chennai 600113, India}
\affiliation{Homi Bhabha National Institute, Anushaktinagar, Mumbai 400 094, India}
\date{\today}
\begin{abstract}
Multicellular organisms exhibit a high degree of structural organization with specific cell types always occurring in characteristic locations. The conventional framework for describing the emergence of such consistent spatial patterns is provided by Wolpert's ``French flag'' paradigm. According to this view, intra-cellular genetic regulatory mechanisms use positional information provided by morphogen concentration gradients to differentially express distinct fates, resulting in a characteristic pattern of differentiated cells. However, recent experiments have shown that suppression of inter-cellular interactions can alter these spatial patterns, suggesting that cell fates are not exclusively determined by the regulation of gene expression by local morphogen concentration. Using an explicit model where adjacent cells communicate by Notch signaling, we provide a mechanistic description of how contact-mediated interactions allow information from the cellular environment to be incorporated into cell fate decisions. Viewing cellular differentiation in terms of trajectories along an epigenetic landscape (as first enunciated by Waddington), our results suggest that the contours of the landscape are moulded differently in a cell position-dependent manner, not only by the global signal provided by the morphogen but also by the local environment via cell-cell interactions. We show that our results are robust with respect to different choices of coupling between the inter-cellular signaling apparatus and the intra-cellular gene regulatory dynamics. Indeed, we show that the broad features can be observed even in abstract spin models. Our work reconciles interaction-mediated self-organized pattern formation with boundary-organized mechanisms involving signals that break symmetry.
\end{abstract}


\maketitle

\section{Introduction}\label{sec1}
\begin{figure}[tbp]%
\centering
\includegraphics[width=0.5\textwidth]{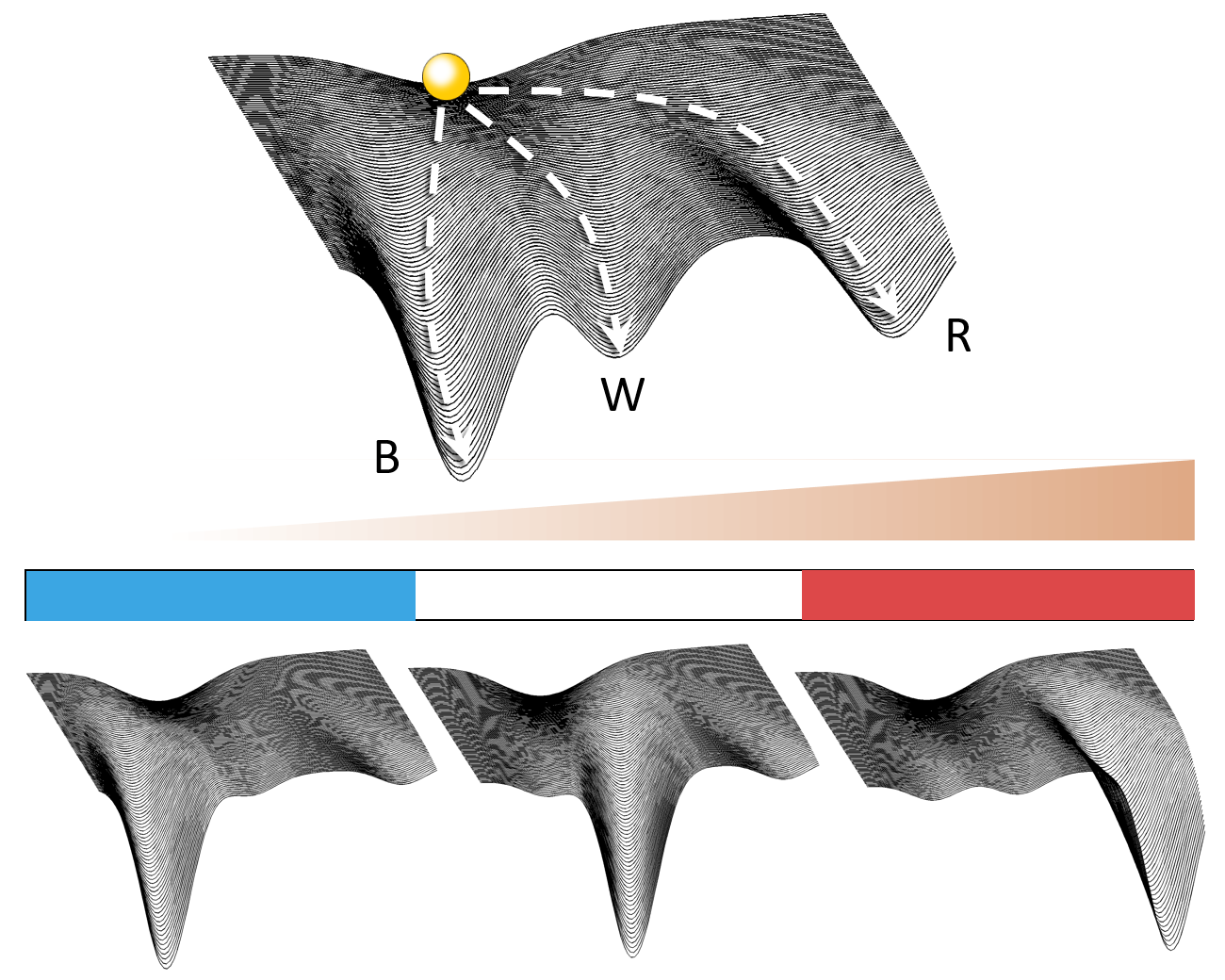}
\caption{\textbf{Convergence to different fates as distinct trajectories 
are followed down the Waddingtonian epigenetic landscape by cells at different locations
in embryonic tissue results in its patterning.}
(Top) The final differentiated state of a cell is given by the expression levels 
of patterning genes B, W and R (say). They correspond to distinct trajectories that
the cell state (represented by the sphere) 
can follow during development depending on its lineage and the contour
of the landscape shaped by both the genetic regulatory interactions intrinsic
to the cell and the environment in which the cell is embedded.
Morphogen molecules diffusing from a source located at one end of the tissue 
can form a concentration gradient (represented as a triangular wedge, center)
providing a key environmental signal that results in differential modulation of
the landscape for cells located at different distances from the morphogen source (bottom).
This resulting bias to trajectories culminating in the specific fates B, W or R, respectively, depending
on the location of the cell, leads to patterning of the tissue (represented by the colors
blue, white and red, respectively) guided by the local morphogen concentration.
The model presented here shows that the fate patterning can also be modulated 
by signaling between neighboring cells, e.g., occurring via contact-mediated
interactions.}
\label{fig0}
\end{figure}

Almost all multicellular organisms possess a characteristic structural organization, whereby
cells possessing identical genetic information differentiate into several distinct types over
the course of development~\cite{Wolpert1991,Wolpert2011,Gilbert2013}. Moreover, such differentiation is ordered spatially, with
specific cell types localized in tissues and organs at particular locations that are almost invariant across individuals. In conjunction with mechanical forces that result in changes in the
geometry of the developing embryo, the acquisition of region-specific fates by cells in
different parts of the organism is responsible for \textit{morphogenesis} - the emergence
of the characteristic body plan of the organism. The key problem that the mechanism underlying
such pattern formation has to solve is to allow a cell to differentiate to the type
that is the most appropriate for 
its spatial location. Thus, it involves relating the processes responsible for a cell acquiring
one of several possible fates (that determine the morphology and function of the 
differentiated cell) with those that allow a cell to obtain
information about its position in the tissue or organ it belongs to.
The process of differentiation can be described from a dynamical perspective as
a trajectory followed by the cell state as it traverses the \textit{epigenetic landscape}
shaped by the genetic regulatory network of the cell and its interaction with
stimuli present in the cellular environment. In this picture, originally proposed by 
Waddington~\cite{Waddington1940}, the possible fates correspond
to different channels in the landscape that the cell state follows, depending upon initial
conditions and perturbations that may arise from internal or external sources (as
shown in Fig.~\ref{fig0}~[top panel], where the cell can acquire any one of three possible fates
B, W or R). A spatial pattern of cell fates will emerge if cells at different locations in 
a tissue can preferentially choose one of the fates over the others based on information about 
their position. This can come about by selective alteration of the landscape for a cell
at a particular position in the tissue, so that the cell state is preferentially
guided towards one of the channels [Fig.~\ref{fig0},~bottom panel].

\begin{figure}[tbp]%
\centering
\includegraphics[width=0.5\textwidth]{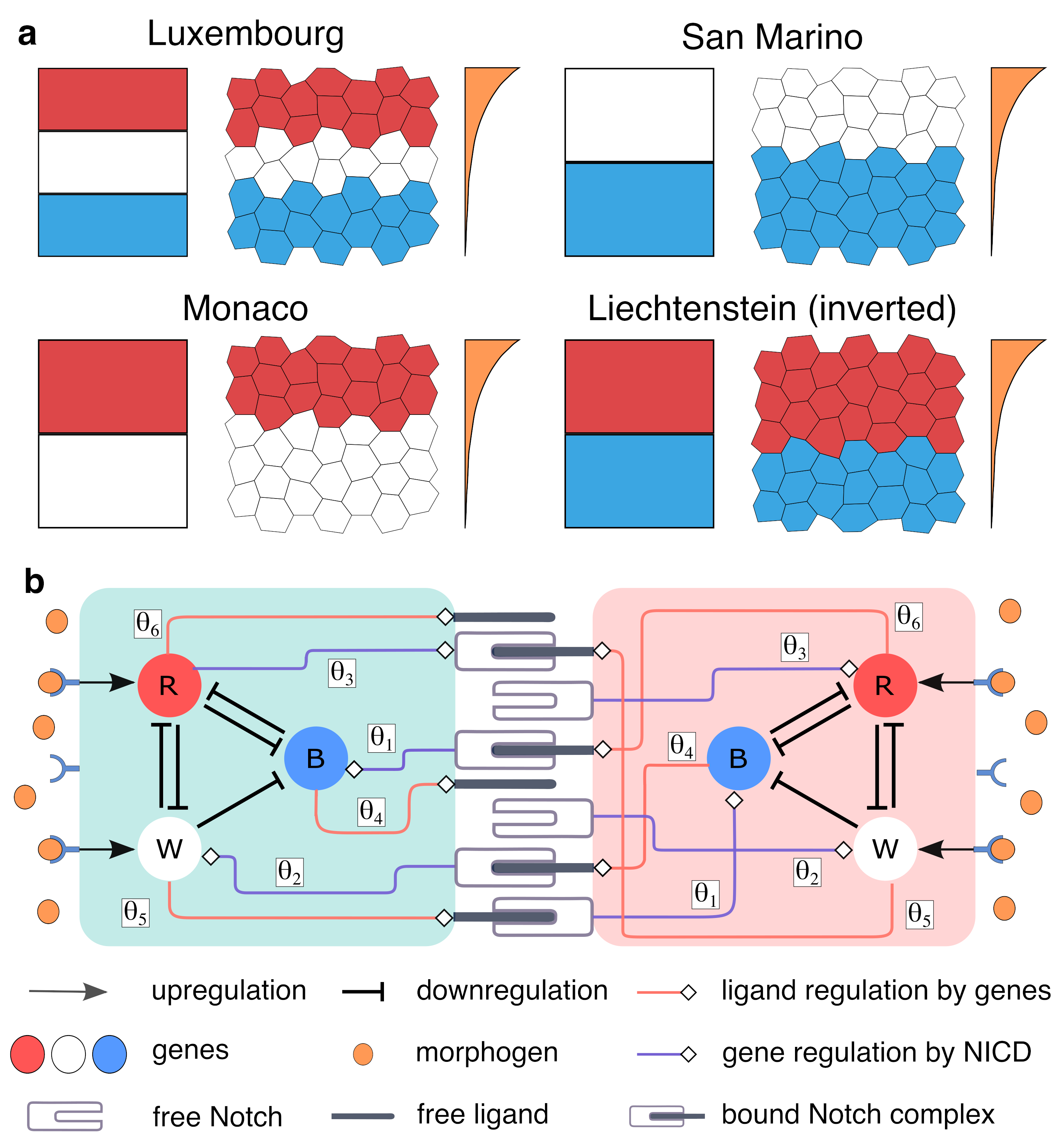}
\caption{\textbf{Morphogen concentration gradients and inter-cellular interactions can jointly determine the spatial pattern of distinct cell fates in embryonic tissue.}
(a)~Flags of small European sovereign states providing an analogous representation
of the qualitatively distinct patterns
that can occur in a cellular assembly (indicated schematically) when each cell
can attain one of three different fates, indicated by the colors blue, white and red,
based on their positional information.
(b)~Schematic representation of contact-mediated interaction occurring via Notch signaling
between a pair of cells responding to morphogen molecules.
The motif comprising mutually repressing patterning genes (B, W and R) functions as
an interpreter of the morphogen concentration to provide positional information.
The expression of the three genes are regulated (with strengths $\theta_{1,2,3}$, respectively) by the Notch intra-cellular domains (NICD) - the product of \textit{trans}-binding
between Notch receptors and ligands. In turn, the gene activities control ligand production
(with respective strengths $\theta_{4,5,6}$).}
\label{fig1}
\end{figure}

Such position-dependent perturbations of the landscape involve symmetry breaking,
which in the context of biological development has typically been associated with molecules collectively referred to as \textit{morphogens}~\cite{Turing1952,Cross1993,Koch1994,Ball1999,Zhang2018}.
The monotonic decrease in the concentration of a morphogen as it diffuses away from a source 
provides a cue to the cells about their relative placement in a tissue~\cite{Crick1970,Driever1988,Teleman2001,Lander2002,Bollenbach2005,England2005,Hornung2005,Lander2007,Ben2010,Yuste2010,Muratov2011}. The resulting deviation from 
spatial homogeneity translates to patterns of differential cell fate expressions.
The \textit{boundary-organized} mechanisms by which the patterns emerge are dependent
on the existence of spatially varying global signals (such as 
morphogen gradients~\cite{Lander2011} or mechanical forces
acting on the tissue~\cite{Kuyyamudi2021_c}).
This is often illustrated using the analogy of a flag (specifically, the French tricolor
flag composed of blue, white and red bands) whose characteristic identity is related to
the relative proportions and sequence of distinct colored regions, independent
of the absolute dimensions of the flag~\cite{Wolpert1969}. Identifying the flag
with a tissue and the colors with distinct fates [Fig.~\ref{fig1}~(a)], 
it is easy to see that the central
question here is how the characteristic partitioning of the domain occurs consistently
into the correct number of fate boundaries, while simultaneously maintaining the right order 
in which the
different cell types appear. Using the ``French flag'' model, Wolpert showed that this
can be solved if (i) the cells have the means to ``infer'' their location in the tissue and (ii)
this allows them to switch on different programs based on the inferred position, leading to distinct fates~\cite{Wolpert1969,Wolpert1989,Sharpe2019}.
Thus, positional information is provided by the local morphogen concentration, which is interpreted
by the intra-cellular gene regulatory network in terms of variations in the expression of 
certain genes (referred to as patterning genes) whose steady-state levels
can be used to represent states corresponding to different fates~\cite{Gurdon2001,Ashe2006,Rogers2011,Gilbert2013}. This
links processes operating in the extra-cellular environment with intra-cellular
gene regulatory dynamics, allowing the Waddington
landscape of cells to be selectively perturbed depending on their position vis-a-vis
the morphogen source (Fig.~\ref{fig0}).

Recent experiments however have suggested that this may not be the whole story.
In particular, work on developmental patterning in the mouse ventral spinal cord
has brought to fore the role played by local cell-cell interactions~\cite{Kong2015}.
Coupling between cells is, of course, known to be the key process underlying
the other important class of pattern formation mechanisms, namely that which relies on 
\textit{self-organization}, as in the reaction-diffusion framework~\cite{Turing1952,Meinhardt1982,Werner2015}.
Therefore, it is intriguing to explore the consequences of possible interplay between
the two principal paradigms proposed for explaining the genesis of biological patterns
in the context of cell-fate patterning in tissues. Here, we do this by 
investigating assemblies of cells that communicate with their neighbors via
contact-mediated signaling, while at the same time being exposed to a  
morphogen concentration gradient. Specifically, we
focus on Notch signaling~\cite{Artavanis1999} as the means by which a cell interacts
with other physically adjacent cells. This involves ligands belonging to its neighbors
binding to the Notch receptors located on the surface of the cell, triggering 
downstream signals
that may eventually affect expression of the patterning genes. Notch has been shown
to be critically important for development in all metazoans~\cite{Artavanis1999,Kopan2009}.
Moreover, it is known that in the presence of noise, such as fluctuations in the 
global signal (morphogen concentration), Notch-mediated interactions can help
the tissue to retain sharpness of the fate boundaries~\cite{Sprinzak2011}, thereby 
enhancing the robustness of developmental dynamics~\cite{Erdmann2009,Lander2011,Lander2013}.
This raises the possibility, explored in detail here, that Notch-mediated inter-cellular interactions 
may modulate the process of morphogen-driven cell-fate patterning to varying extents, giving
rise to ``flags'' that may deviate quite markedly from the (tricolor) pattern one would
expect in the absence of such interactions [Fig.~\ref{fig1}~(a)].

In this paper we show that contact-mediated interactions between cells
can provide a mechanism by which the cell can use information collected from
its neighborhood to adapt its fate decision that would otherwise solely be guided by global
positioning information, e.g., as obtained from the local concentration of
a morphogen gradient. Our theoretical demonstration is 
consistent with reported results of experiments performed on the ventral spinal cord
of mice which showed that the proportion of regions exhibiting
distinct cell fates differ for the situation when Notch inter-cellular signaling is absent as compared
to when it is present~\cite{Kong2015}. Our model incorporates an explicit connection between the 
genes that respond to the morphogen signal and Notch signaling that 
allows physically adjacent cells to communicate. Specifically, the downstream effector of
the Notching signaling pathway is assumed to regulate patterning gene expression,
which in turn controls the production of the ligands that bind to
the Notch receptors. The interaction between the patterning genes is described
by equations originally proposed in the context of patterning in the
vertebrate neural tube where a circuit comprising three mutually regulating genes 
serves as the interpreter for the positional information encoded in the concentration gradient of
the Sonic Hedgehog (Shh) morphogen~\cite{Dessaud2008,Dessaud2010,Balaskas2012}.

We systematically vary the nature of each of the regulatory links between the 
actors involved in inter-cellular interaction and those responsible for local cell fate decisions,
and identify the ones that are most instrumental in altering the cell fate pattern.
In particular, we show that upregulation of patterning gene expression by the
Notch downstream effector consistently results in large deviation of the pattern from
that observed in the absence of Notch-mediated coupling. Large perturbations
in the sizes of the regions occupied by the various types of differentiated cells 
and the sequence in which they are arranged can potentially reduce the viability of (or otherwise
be disadvantageous to) an organism in terms of its survival. This suggests the
presence of strong evolutionary constraints on the nature of any regulatory
interactions between the inter-cellular signaling machinery and the fate determining genes.
We show that our results are not tied to the specific choices we have made in our model
by demonstrating qualitatively identical results in a model variant where the
Notch directly regulates ligand production, as well as, the patterning gene expression
(the latter not affecting the ligands). Indeed, even generic models such
as that of binary spins interacting via nearest-neighbor exchange interactions
can illustrate the broad features of how local interactions can modulate the
collective behavior of a system responding to a global field. Thus, 
they allow intermediate-scale phenomena (spanning the cellular neighborhood)
to modulate cellular-scale fate decisions that would otherwise be determined solely
by tissue-scale signals (the morphogen concentration gradient). In terms of the
Waddington framework discussed above, the contact-mediated inter-cellular communication
provides an alternative mechanism by which to mould the topography of the
landscape shaped by the morphogen signal, that allows subtle (and not so subtle)
variations in the flag formed by cell-fate decisions at the scale of the entire
cellular assembly.

\section{Methods}\label{sec2}

\begin{table*}[htbp]
\small
\begin{center}
\begin{tabular}{|c|c|c|c|c|c|c|c|c|c|c|c|c|c|c|c|c|c|c|}\hline
$S_M(0)$&$\lambda_M$ &$\alpha$&$\beta$&$\gamma$& $\beta_L$&$\beta_{N^b}$& $K$ & $K_N$ & $k_1$ & $k_2$ & $k_3$ & $\tau_L$ & $\tau_{N^b}$ &$h_1$&$h_2$&$h_3$&$h_4$&$h_5$\\\hline
100&0.3&4&6.3&5&5&5&1&1&1&1&1&1&1&6&2&5&1&1\\\hline
\end{tabular}
\caption{The values (second row) for the model parameters (first row) used for the simulation results reported here.}\label{table1}
\end{center}
\end{table*}

We consider a $1$-dimensional cellular array of length $L$ placed in a morphogen concentration gradient
whose source is assumed to be located at one end of the array. Each cell responds to the
local density of morphogen molecules with which its receptors (located on the cell surface) can bind.
The morphogen concentration $M$ is assumed to be the outcome of a synthesis-diffusion-degradation (SDD) model having the same mean lifetime for the molecules across space. This yields an
exponentially decaying profile for $M$ (with the maximum located at the source) in the
steady state~\cite{Crick1970}, viz.,
$M(n) = M_0\,\exp(-\lambda_{M}n)$, where the integer index $n$ varies over the range $[0,L)$.
The cellular response to binding with morphogen can be measured in terms of the
concentration $S_M$ of downstream signaling molecules that are triggered upon successful binding.
Assuming that the cellular response mirrors the external morphogen concentration on average,
we can express the spatial variation of the response as $S_M (x) = S_M (0) {\rm exp} (-x/\lambda_M)$, with $x$ representing the distance from the morphogen source.
As indicated in Table~\ref{table1}, we have
chosen for all our simulations $S_0 = 100$ and $\lambda_M = 0.3$, and
have verified that our results are not qualitatively sensitive to changes in these values.

Each cell possesses a morphogen interpretation module comprising a set of genes that can regulate
each other's activity and whose expression levels are modulated by the signaling
molecules downstream of the receptors triggered by the morphogen.
The specific gene circuit that we have chosen for our simulations is composed of three
genes B, W and R (which we refer to as patterning genes), using a model that has been
used to describe the emergence of tissue differentiation in the vertebrate neural tube's
ventral region~\cite{Balaskas2012}. The relevant morphogen in this case is Sonic hedgehog (Shh),
while the genes are Pax6, Olig2 and Nkx2.2, with Pax6 being expressed even when the morphogen
is not present and hence can be identified as the pre-patterning gene (taken to be gene B
according to our naming convention). In the initial stage, before the morphogen gradient
makes itself felt fully by the cells, this gene will have a higher level of expression compared
to the other two genes. However, in the steady state, the morphogen, by promoting
the activity of W and R genes, may induce repression of B in parts of the array.

The genes W and R mutually repress each other, as do the genes R and B; however, while W can
repress B, B has no effect on W [see Fig.~\ref{fig1}~(b)]. We assume that the fate of
each cell is decided by the gene that has the highest level of expression in it in the steady
state following initial transient dynamics. Thus, regions indicated by the colors blue, white and red,
correspond to cells where the genes B, W and R are expressed most strongly, respectively.
The following equations describe the time-evolution of the gene expressions:
\begin{align}
  \label{eq1}
  \frac{dB}{dt} &=
  \frac{\alpha + \varphi_{1}\left.\frac{N^b}{K_N}\right.}{1 +
  \left(\frac{R}{K}\right)^{h_1} + \left(\frac{W}{K}\right)^{h_2} +
  \xi_{1}\left.\frac{N^b}{K_N}\right.} - k_{1}B\,,\\
   \label{eq2}
   \frac{dW}{dt} &=
   \frac{\beta S_M + \varphi_{2} \left.\frac{N^b}{K_N}\right.}{1 + S_M +
   \xi_{2}\left.\frac{N^b}{K_N}\right.}~
   \frac{1}{1 + \left(\frac{R}{K}\right)^{h_3}} - k_{2}W\,,\\
   \label{eq3}
   \frac{dR}{dt} &=
   \frac{\gamma S_M + \varphi_{3} \left.\frac{N^b}{K_N}\right.}{1 + S_M +
   \xi_{3}\left.\frac{N^b}{K_N}\right.}~
   \frac{1}{1 + \left(\frac{B}{K}\right)^{h_4} +
   \left(\frac{W}{K}\right)^{h_5}} - k_{3}{R}\,,
\end{align}
where the maximal growth rates and decay rates for expression for each of the genes
are represented by the
parameters $\alpha, \beta , \gamma$ and $k_1,k_2,k_3$, respectively.
The response functions are specified by the parameters $K$, $K_N$ and $h_1,h_2,h_3,h_4,h_5$.

We also consider contact-mediated inter-cellular interactions via the Notch signaling pathway
~\cite{Artavanis1999,Kopan2009}. This is incorporated into the expression dynamics of the
genes above by the parameters $\varphi_1,\varphi_2,\varphi_3$ and $\xi_1,\xi_2,\xi_3$.
The effect of the inter-cellular signaling on the dynamics of the system can be described by
augmenting the above equations with those describing the time evolution of concentration
of Notch ligand $L$ and the Notch intra-cellular domain (NICD) $N^b$, viz.,
\begin{align}
\label{eqL}
   \frac{dL}{dt} &=
   \frac{\beta_L + \phi_{4}\frac{B}{K} + \phi_{5}\frac{W}{K} + \phi_{6}\frac{R}{K}}{1
   + \zeta_{4}\frac{B}{K} + \zeta_{5}\frac{W}{K} + \zeta_{6}\frac{R}{K}}
   - \frac{L}{\tau_L}\,,\\
\label{eqN}
   \frac{dN^b}{dt} &=
   \frac{\beta_{N^b}L^{trans}}{K + L^{trans}} - \frac{N^b}{\tau_{N^b}}\,.
\end{align}
The maximum growth rates of the ligand and the NICD are given by $\beta_L,\beta_{N^b}$,
while their mean lifetimes are represented by $\tau_L,\tau_{N^b}$, respectively.
Upon the binding of ligands $L^{trans}$ of a neighboring cell to a cell's surface Notch receptors,
the intracellular domain of the receptor is released and it subsequently translocates
itself to the cell nucleus. We have assumed here a sufficiently high density of receptors
for each cell so that they are not saturated.
A key feature of our modeling is that the Notch signaling machinery and the morphogen
interpretation module are considered to be able to control each other [Fig.~\ref{fig1}~(b)].

The ligand can be activated, inhibited or not affected at all by each of the patterning genes
(promotion/repression being analogous to the situation corresponding to Jagged and Delta ligands,
respectively~\cite{Shimojo2011,Manderfield2012,Boareto2015})
while the genes themselves can again be either regulated by NICD in a positive or negative
manner or unaffected [Fig.~\ref{fig2}~(a)].
Thus, depending on whether the regulation occurs at all and if so, then depending on its nature,
there are $3^6 = 729$ classes of inter-cellular coupling (which includes also the trivial uncoupled
case). If a patterning gene is upregulated by NICD, the corresponding parameters
$(\varphi_{i},\xi_{i})$ are given by $(\theta_{i},1)$, while in case of downregulation
they are given by $(0,\theta_{i})$ (with $i=1,2,3$ labeling the three genes).
Similarly promotion of the ligand by a patterning gene will be represented by the
corresponding parameters $(\phi_{j},\zeta_{j})$ adopting the values
$(\theta_{j},1)$, while repression corresponds to the parameters having the values
$(0,\theta_{j})$ (the three genes indexed as $j=4,5,6$, respectively). For both types of interactions, absence of regulation of/by a gene
will correspond to both the corresponding parameters having the values $0$.

The values of the model parameters  (shown in Table~\ref{table1}) are chosen such that
in the absence of inter-cellular interactions (viz., $\varphi_i =0$ and $\xi_i= 0$, $\forall i$)
we obtain a pattern that corresponds to the three fate segments having
equal length and in the correct chromatic order (B,W,R).
In order to investigate how the coupling between cells regulates the pattern,
for each of the $729$ possible types of connections
between the Notch signaling apparatus and the patterning genes we simulate the
system dynamics with $10^4$ distinct combinations of the coupling
strengths $\theta_1, \ldots, \theta_6$ sampled randomly over an uniform distribution
within the range $[1,10]$ for up-regulation and within $[0.1,1]$ for down-regulation.

\section{Results}\label{sec3}
\begin{figure}[tbp]%
\centering
\includegraphics[width=0.5\textwidth]{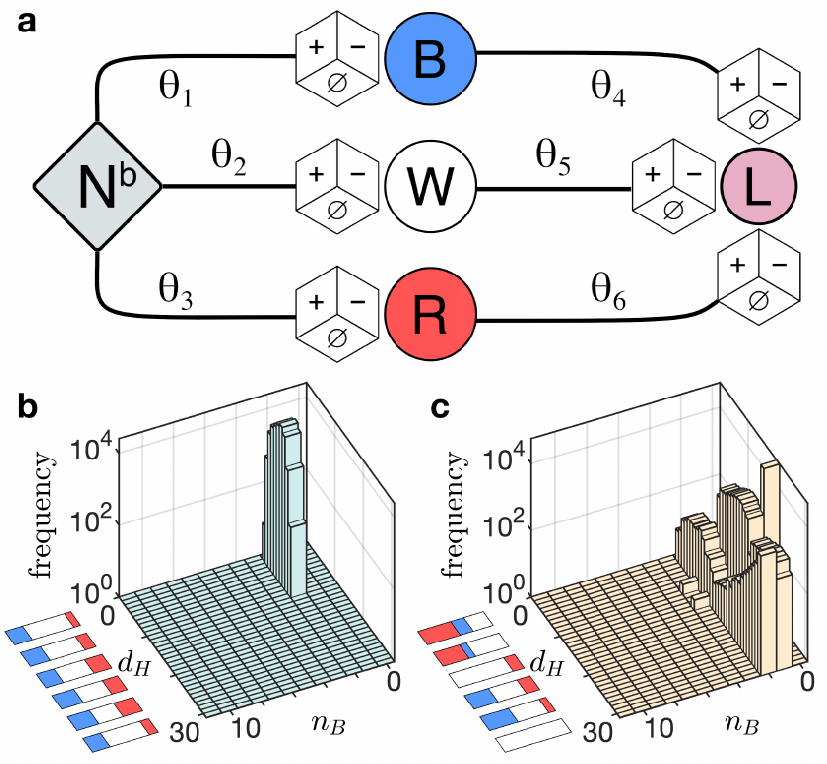}
\caption{\textbf{The nature of interactions between the patterning genes and the components
of inter-cellular Notch signaling shapes the spatial patterns of cell fates qualitatively,
as well as, quantitatively.}
(a)~Motif indicating the different ways in which the interactions can be classified,
depending on how the NICD regulates each of the genes and how the genes, in turn, affect
ligand production ($+$: upregulation, $0$: no effect, $-$:~downregulation). The strength
of each link is indicated by $\theta_i$ ($i=1,\ldots,6$).
(b-c) Representative frequency distributions of patterns obtained by two possible classes
of interactions, constructed by randomly sampling values of $\theta$ for a one-dimensional
domain consisting of $30$ cells. Each pattern occupies a specific position in the
space defined by the observables $n_B$ (number of boundaries between regions
corresponding to distinct fates) and $d_H$ (the bit-wise distance between the pattern
and the idealized flag having equal divisions of B, W and R).
Sample flags obtained for each type of interaction are shown along the axis
representing $d_H$.
The interaction motif $(-,0, -, -, -, +)$
corresponding to panel (b) produces flags close to the idealized template,
while that corresponding to panel (c), viz., $(+,+,+,+,0,-)$,
produces some of the most divergent patterns. This motif notation designates the nature
of each of the interactions in the same sequence $i=1,\ldots,6$ as that in which their
strengths are represented by $\theta_i$ ($i=1,\ldots,6$).
}\label{fig2}
\end{figure}
In order to characterize the various patterns that arise in the presence of inter-cellular coupling
as we alter the qualitative and quantitative nature of the interactions, we note that the
regions exhibiting different cell fates may not only differ in terms of their size (lateral extent)
but also the sequence in which they occur in the domain, and even the number of times
that a contiguous region with a particular fate appears in the flag. In order to take into
account these distinctions quantitatively we employ two different measures, viz.,
(i) the total number of boundaries $n_B$ between regions having distinct fates and (ii) a
metric for the difference between the observed pattern and the flag obtained in the
absence of coupling, that measures the binary distance between the fate in each
cell position in the
two cases (i.e., $=0$ if they are identical, and $=1$ otherwise) and then sums over all positions.  This latter is identical to a Hamming distance
between two symbolic strings and hence represented as $d_H$.
As seen from Fig.~\ref{fig2}~(b-c), depending on the motif being considered we can obtain
flags that can be quite close to the idealized one having equal sized segments of B, W and R
(in that order) [e.g., panel (b) where NICD downregulates B and R but does not affect W, while
ligand production is downregulated by B and W but upregulated by R],
or extremely divergent patterns [e.g., in panel (c) for the case where NICD upregulates
all genes, while B upregulates, R downregulates and W does not affect ligand production].

\begin{figure}[tbp]%
\centering
\includegraphics[width=0.49\textwidth]{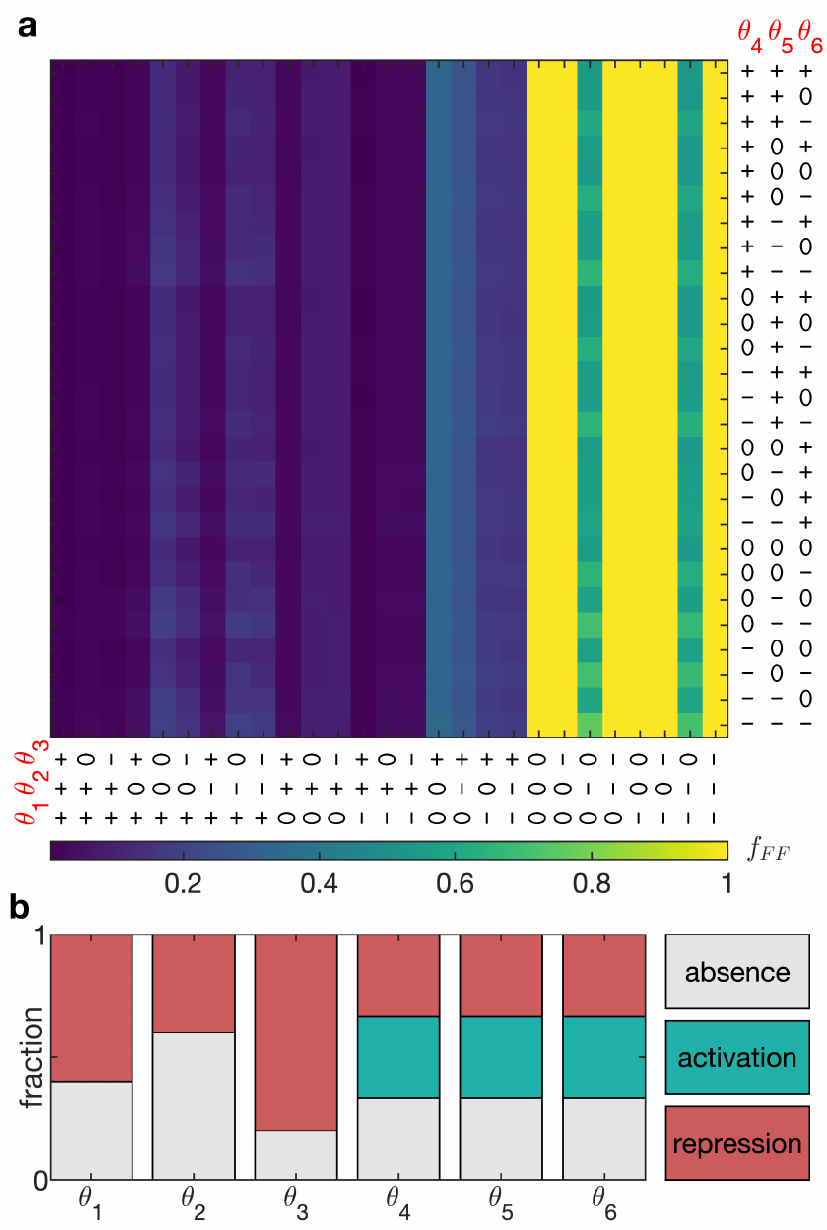}
\caption{\textbf{Upregulation of any patterning gene by Notch signal
yields distorted patterns that do not conserve the number of chromatic regions or their sequential order seen in absence of inter-cellular interactions.}
(a) Matrix displaying normalized frequencies $f_{FF}$ of ``French'' flags 
(having exactly $3$ chromatic regions occurring in the order B,R,W)
obtained for each of the $3^6 (= 729)$ possible interaction motifs. The frequency for each motif is estimated from $10^4$ realizations
with randomly sampled initial conditions and values of the parameters ($\theta_1, \ldots, \theta_6$). Each of the $3^3$ columns (rows) correspond to a specific combination of up/down/no regulation [$+/-/0$] of the patterning genes by NICD (of ligand production by the patterning genes).
(b) The fractions of the three different types of regulation, viz., $0$:~absence of regulation, $+$: activation, $-$: repression, yielding ``French'' flags for each of the $6$ interactions (whose
strengths are given by $\theta_1, \ldots, \theta_6$).
Note that, upregulation of the chromatic genes by NICD never lead to such flags.}
\label{fig3}
\end{figure}
In order to systematically evaluate the patterns resulting from each of the $3^6$ possible
interaction motifs, we quantify the fraction $f_{FF}$ of realizations (with randomly sampled
parameters $\theta_1, \ldots, \theta_6$) for each motif that gives
rise to a ``French'' flag, i.e., a pattern characterized by $n_B=2$ fate boundaries
and the correct chromatic sequence of B, W and R.
This is because flags that do not conserve $n_B$ or the chromatic order
of the idealized flag represent marked aberrations that may be undesirable in the
context of tissue development.
Fig.~\ref{fig3}~(a) shows a matrix of $f_{FF}$ for all the motifs, estimated from
$10^4$ realizations in each case. It is immediately clear that the nature of
regulation of ligand production by the patterning genes plays an extremely minor
role (if at all) in regulating the cell fate pattern, as is evident from the relative lack of
variation in $f_{FF}$ along each column (the different rows correspond to
different choices, viz., up/down/no regulation of ligand production by the patterning genes).
Further, we note that if the NICD upregulates any of the patterning genes, the resulting
pattern almost never resembles a ``French'' flag. Thus, it appears that in order
to yield patterns close to that obtained in the absence of any inter-cellular interactions,
the NICD should either downregulate or not affect the genes B, W and R. We would
like to point out however that the downregulation of the W gene, either in absence
of regulation of the other genes or downregulation of only the B gene by NICD,
can result in at least some fraction of the flags departing markedly from the
idealized pattern.

The importance of the different interactions in regulating the fate pattern can be
seen clearly from Fig.~\ref{fig3}~(b) which shows
the fractions of
each type of regulation (up/down/none) that give rise to a ``French'' flag
for each of the 6 regulatory
connections (whose strengths are represented by $\theta_1, \ldots, \theta_6$, respectively)
linking the pattern gene expression dynamics to Notch signaling.
Consistent with the results stated above, we find that  the nature of regulation
is irrelevant for the interactions
whose strengths are $\theta_4,\theta_5,\theta_6$ (i.e., the regulation of the
ligand production by the 3 patterning genes),
as each of the three possible types are equally likely to generate such
a pattern. However, for the regulatory interaction of the patterning genes by NICD,
only upregulation or no regulation can give rise to a ``French'' flag. This also resonates
with our earlier results on the sensitivity of the pattern to variation in the values of the different
parameters using either Sobol variation-based analysis or Sloppy model analysis
techniques~\cite{Kuyyamudi2021}. These results had suggested that the
parameters $\theta_2$ and $\theta_3$ play the most important role in determining
the modulation of the cell fate pattern by the inter-cellular interaction.
\begin{figure}[tbp]%
\centering
\includegraphics[width=0.5\textwidth]{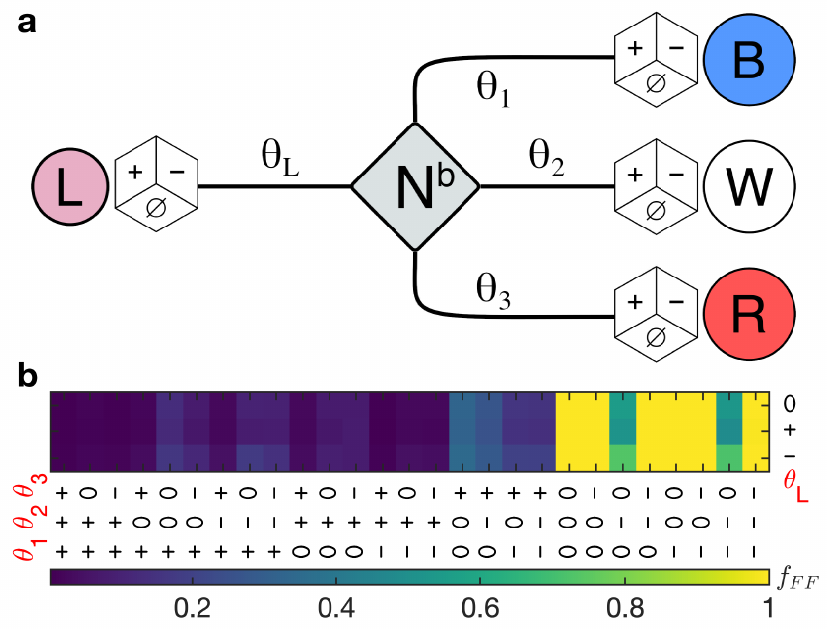}
\caption{\textbf{An alternative framework for achieving distinct cell fates through
Notch-mediated cell-cell coupling yields qualitatively similar patterning behavior in the system,
underlining its robustness.}
(a)~Motif representing a different scheme for connecting intra-cellular interaction
via Notch signaling to patterning gene expression. In contrast to the framework discussed earlier,
here the NICD regulates the ligand production as well as the three genes B, W and R, while
the genes do not affect the ligand. The nature of regulation by NICD can be one of three
types ($+$: upregulation, $0$: no effect, $-$:~downregulation) for each of the genes and the
ligand, the strengths of the links being indicated by $\theta_{1,2,3}$ and $\theta_L$, respectively.
(b) Matrix displaying the normalized frequency $f_{FF}$ of ``French'' flags having $n_B=2$ boundaries with
the same chromatic order as the idealized flag obtained for each of the $3^4 (= 81)$ possible interaction motifs. The frequency for each motif is obtained from $10^4$ realizations of the model
with randomly sampled initial conditions and values of the parameters ($\theta_{1,2,3,L}$). Each of the $3^3$ columns correspond to a specific combination of up/down/no regulation ($+/-/0$)
of the patterning genes by NICD, while the $3$ rows represent up/down/no regulation ($+/-/0$) of ligand production by the NICD.
}
\label{fig4}
\end{figure}

We would like to note that, for lateral induction or inhibition effected by inter-cellular
signaling, it is known from experiments that Notch directly regulates the concentration
of its ligand~\cite{Collier1996,Sprinzak2010,Sprinzak2011}. Here, in order to connect inter-cellular signaling
to gene expression leading to fate determination, we have considered a setting in
which this regulation takes place indirectly via the action of NICD on the patterning genes,
and subsequently, that of the genes on ligand production. However, we can also
consider a different way by which patterns of distinct cell fates can be
influenced by inter-cellular Notch interactions.
In this alternative approach, we choose NICD to directly
regulate the ligand production, as well as, the three genes in
any of three possible ways (up/down/no regulation) [Fig.~\ref{fig4}~(a)].
In contrast to the preceding model, the gene expression does not affect the
components of Notch signaling.
Almost all the equations describing the system dynamics earlier remain
unchanged (Eqs.~\ref{eq1},~\ref{eq2},~\ref{eq3} and ~\ref{eqN}) in this alternative model,
except the equation describing the time-evolution of the concentration
of Notch ligand $L$, viz.,
\begin{align}
   \frac{dL}{dt} &=
   \frac{\beta_L + \phi_{L} N^b}{1  + \zeta_{L} N^b}
   - \frac{L}{\tau_L}\,,
\end{align}
where the parameters ($\phi_L,\zeta_L$) are given by ($\theta_L,1$) in the case
of upregulation of ligand by the NICD, while for downregulation they are
($0,\theta_L$). If there is no regulation, both parameters are set to $0$.
Fig.~\ref{fig4}~(b) shows the fraction $f_{FF}$ of  ``French'' flags for all
possible $3^4 (= 81)$ qualitatively distinct interaction motifs that are allowed
in this new framework. As before these fractions are estimated from $10^4$
realizations in each case using randomly sampled initial conditions and values
of the parameters $\theta_{1,2,3,L}$. We immediately note the high degree
of similarity of the results with those of the earlier model [compare the
columns of the matrix in Fig.~\ref{fig3}~(a) and that in Fig.~\ref{fig4}~(b)].
Thus, upregulation of the patterning genes by NICD always results in the pattern diverging
markedly from the idealized one having $2$ boundaries and the chromatic sequence B,W,R,
while the manner in which ligand production is regulated seems to have little effect
on the pattern. The near identity of the patterning behavior for the two coupling
frameworks we have considered imply that the role played by inter-cellular interactions
in determining morphogen-driven tissue patterning is robust to variations in the precise
details of the mechanism through which Notch signaling and the gene expression governing
cell fates are related.

\begin{figure}[tbp]%
\centering
\includegraphics[width=0.5\textwidth]{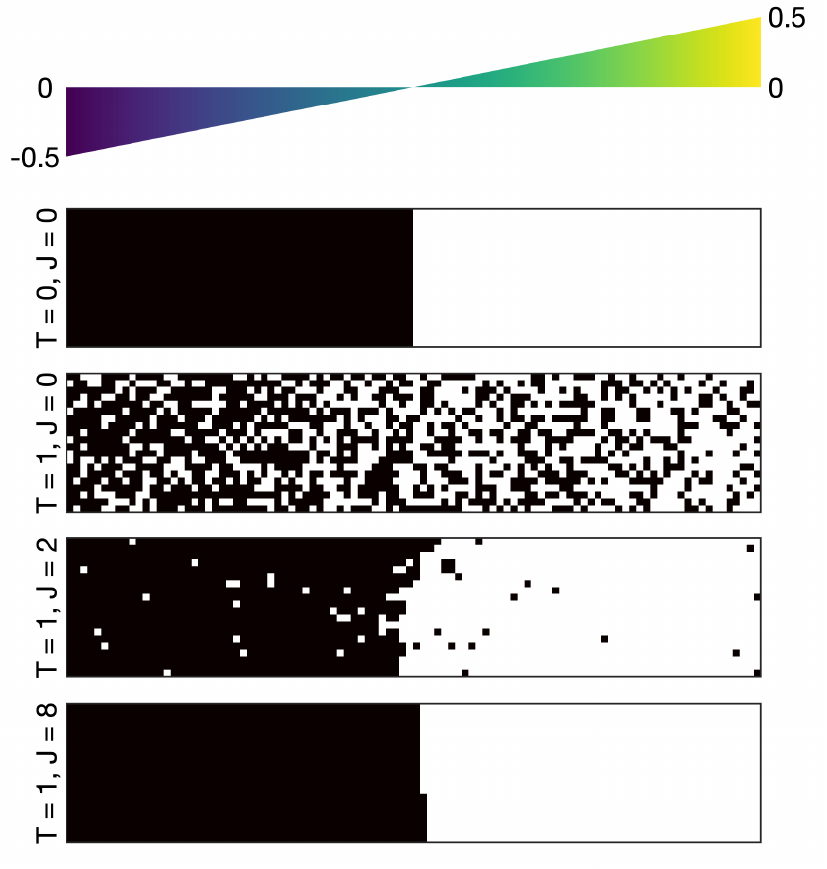}
\caption{\textbf{Local interactions can reinforce a pattern guided by a global field
in the presence of noise.}
A binary spin model representation of the collective behavior emerging as a result
of a system being subject to both a spatially varying external field $H$ ranging
from$-1/2$ to $+1/2$ (top), and
interaction $J$ between neighboring elements that favor parallel orientation
between them (i.e., ferromagnetic). The panels show the pattern of spin orientations
(down: black, up: white) under different conditions of temperature $T$ and strength
$J$ of exchange interactions between each spin with its $4$ nearest neighbors.
The displayed state in each panel is
obtained for a system comprising $20 \times 100 $ spins, resulting after $10^6$ steps
of a Metropolis algorithm starting from random initial conditions.
}
\label{fig5}
\end{figure}

\section{Discussion}\label{sec4}
Our results reported here show how local interactions between neighboring elements
(mediated by Notch signaling)
can modulate the emergent response of the system to a global signal, specifically,
a spatially varying external field (set up by a diffusing morphogen).
Such phenomena can arise in contexts far removed from that of cell fate patterning
in tissues that we use here to motivate the problem. In particular, one can use
the generic Ising model used to study collective ordering in arrays of binary state
elements (represented as spins that can either be in ``up'' or ``down'' orientations).
The analog of inter-cellular interactions in this case is the exchange interaction
that couples the state of a spin with those of its neighbors - the neighborhood being
specified by the geometry of the lattice being considered. Similarly, the concentration
gradient of the morphogen is echoed by a magnetic field whose intensity varies
monotonically over space. One can, in principle, also consider temperature that
introduces thermal fluctuations which disrupt the pattern imposed by the field.
In the tissue, such stochastic effects will arise upon considering the presence of
intrinsic and extrinsic noise affecting the system dynamics. While this has not
been investigated here, we have elsewhere~\cite{Chandrashekar2022}
looked at how inter-cellular interactions can
make a pattern robust against noise using a simpler setting of a single boundary
separating regions with two distinct fates.

Fig.~\ref{fig5} shows the outcome of evolving a finite size $2$-dimensional lattice of Ising
spins for $500$ Monte Carlo (MC) steps,
where the spins interact with their
neighbors in the four cardinal directions (spins located at the
boundaries interact with fewer neighbors than those in the bulk).
The steady state behavior of the system for the case where only the field is present
(first row: $T = 0$, $J = 0$) corresponds to the idealized flag that we
observe in the tissue model in absence of Notch signaling. We subsequently
increase the temperature to a finite value ($T=1$) and investigate the resulting
pattern in absence of local interactions (second row: $J = 0$)
and for the cases where the interactions are relatively weak (third row: $J = 2$)
and when they are strong (fourth row: $J=8$). We observe that while noise
completely distorts the pattern resulting from the weak applied field, introducing
strong spin-spin interactions restores the pattern seen in absence of thermal fluctuations
and exchange interactions - analogous to results in the case of tissue patterning~\cite{Chandrashekar2022}. Thus, it is possible to connect the specifically biological problem
that motivated our study with physical analogues that may be easier to understand
analytically.

To put our modeling framework in a broader context, we return once again to the analogy of the French flag that was used by 
Wolpert to discuss the key problem
of differential fate expression in cells according to their spatial location leading to 
consistent tissue patterning. One of the central features emphasized by Wolpert
is the size invariance of the pattern: regardless of the scale at which the tricolor
is represented - be it on a lapel pin or displayed across the side of a building - the French flag is always
recognizable by the characteristic sequence of blue, white and red domains of equal width.
In biological tissue, this will correspond to the same proportion of cells converging
to different fates regardless of the absolute size of the domain over which the morphogen
gradient is imposed. From a conventional perspective, where the attractor (corresponding
to the differentiated state) to which a cell converges to is a function of the local morphogen
concentration, this may appear somewhat difficult to explain unless the gradient itself
adapts to the dimension of the domain. 

In the absence of such adaptation, it would appear
that increasing or decreasing the size of the tissue would result in distortion of the
pattern as the size of the regions corresponding to different fates will no longer be
proportional. In order for the original pattern to be reproduced in domains having
different sizes, it would be necessary for a cell at a particular location to be aware of
the size of the domain it is part of - a problem analogous to that encountered in 
quorum sensing~\cite{Miller2001,Waters2005} - so that it can appropriately adjust the information
provided by the local morphogen signal. Thus, in a smaller domain, the cells located in
the region farthest away from the morphogen source should express the fate B consistent
with low morphogen signal, even though in absolute terms the concentration of morphogen
they may be detecting would have given rise to the fates W or R in a larger domain.

To get a glimpse of how a cell can possibly gather the necessary information to be able to perform
this recalibration, we note an analogous situation in the retina which needs to maintain
a high level of sensitivity to the optical signal it receives under a broad range of varying light intensity. A response curve that varies gradually over the entire range of intensity would
have extremely low contrast while one which changes sharply over a narrow range of 
intensities would be insensitive to variation over most of the full intensity range~\cite{Blanchard1918,Werblin1973}.
This problem is solved by the cells adaptively shifting the response curve according to
the mean intensity of the signal received by neighboring cells, thereby achieving
both sensitivity and contrast. We suggest that the problem of maintaining scale
invariance of cell fate pattern in tissues can possibly also be resolved in a similar manner,
with a cell acquiring knowledge of the larger scheme of things by exchanging information
with its neighbors, using either Notch signaling or other contact-mediated communication
mechanisms.
This highlights the key theme that we explore here, viz., integrating the two principal
classes of pattern formation mechanisms operating in biology~\cite{Lander2011}, 
one involving a global signal providing position information (the boundary-organized patterning paradigm) and the other using local interactions between the elements (the self-organized
paradigm), can yield novel insights in the quest to understand how form and organization 
arises during the development of an organism~\cite{Green2015,Kuyyamudi2021_b}.

\section{Conclusion}\label{sec4}
Cellular differentiation has been often described in terms of dynamical trajectories on
an epigenetic landscape such that the cell state eventually converges to any one of multiple
attractors that correspond to distinct cell fates~\cite{Waddington1957,Ferrell2012,Wang2014,Mojtahedi2016,Moris2016}.
For a single cell, the asymptotic behavior
resulting from this dynamics - i.e., the fate it attains - is decided by the initial condition
and external environment. However, when we consider the problem of tissue patterning
wherein the collective behavior of many neighboring cells are at play in deciding their
fates, additional influences need to be considered. In particular, inter-cellular interactions
can modify the landscape over which the state trajectory of individual cells evolve.
To demonstrate this, we chose a relatively simple model involving intra-cellular dynamics
that results in the cells choosing between one of three different fates based upon
the morphogen concentration (the external environment) and contact-mediated signaling
with neighboring cells, which is realized in our model in terms of Notch signaling.
While it is possible to couple the behavior of the patterning genes responsible for fate
choice and the inter-cellular signaling in multiple ways, we show that two different ways
of having the Notch signal affect the gene expression yields qualitatively identical
results, suggesting that the broad contours of the behavior reported here are
insensitive to the specific details of the model. The key finding is that cell-cell interactions
can indeed modify the cell fate pattern imposed by the field realized by a
gradient of diffusing morphogen molecules. Thus, while morphogens have been
seen as the determinant of the broad contours in which an organism's body is organized,
the signaling between cells provide a flexibility that can allow them to adapt their
eventual fates to local information rather than be enslaved to the global commands
issued by morphogens. The analogy with collective ordering in abstract models such
as Ising spin lattices suggest that this is a general feature, and possibly is at work
in developmental pattern formation at multiple scales and using different coupling mechanisms.

\begin{acknowledgments}
We would like to thank Marcin Zag\'{o}rski for helpful discussions.
SNM has been supported by the IMSc Complex Systems Project (12th
Plan), and the Center of Excellence in Complex Systems and Data
Science, both funded by the Department of Atomic Energy, Government of
India. The simulations required for this work were
supported by IMSc High Performance Computing facility (hpc.imsc.res.in) [Nandadevi].
\end{acknowledgments}



\end{document}